# Towards Ethical AI in Power Electronics: How Engineering Practice and Roles Must Adapt

**Fanfan Lin**, *Zhejiang University-University of Illinois Urbana-Champaign Institute*, **Peter Wilson**, *University of Bath*, **Xinze Li**, *University of Arkansas*, **Alan Mantooth**, *University of Arkansas*

**Abstract:**
Artificial intelligence (AI) is rapidly transforming power electronics, with AI-related publications in IEEE Power Electronics Society selected journals increasing more than fourfold from 2020 to 2025. However, the ethical dimensions of this transformation have received limited attention. This article underscores the urgent need for an ethical framework to guide responsible AI integration in power electronics, not only to prevent AI-related incidents but also to comply with legal and regulatory responsibilities. In this context, this article identifies four core pillars of AI ethics in power electronics: Security & Safety, Explainability & Transparency, Energy Sustainability, and Evolving Roles of Engineers. Each pillar is supported by practical and actionable insights to ensure that ethical principles are embedded in algorithm design, system deployment, and the preparation of an AI-ready engineering workforce. The authors advocate for power electronics engineers to lead the ethical discourse, given their deep technical understanding of both AI systems and power conversion technologies. The paper concludes by calling on the IEEE Power Electronics Society to spearhead the establishment of ethical standards, talent development initiatives, and best practices that ensure AI innovations are not only technically advanced but also oriented toward human and societal benefit.

Interest in artificial intelligence (AI) within the power electronics community has surged in recent years. A search across the IEEE Power Electronics Society (PELS) portfolio, including *IEEE Journal of Emerging and Selected Topics in Power Electronics* (JESTPE), *IEEE Transactions on Power Electronics* (TPEL), and *IEEE Power Electronics Magazine*, shows that the number of AI-related papers published between 2020 and 2025 has increased around fourfold, as listed in Table 1. In addition, tutorials and special sessions on AI have been featured at major conferences such as the *Applied Power Electronics Conference* (APEC 2025) and *IEEE Energy Conversion Conference and Expo* (ECCE 2025), further demonstrating the community's growing interest in this field.

**TABLE 1: Yearly Counts of AI-Related Papers in Selected IEEE PELS Journals and Magazine (Year 2020–2025)**

| Publication / Year | 2020 | 2021 | 2022 | 2023 | 2024 | 2025 | Growth (2020–2025) |
|---|---|---|---|---|---|---|---|
| *JESTPE* | 5 | 7 | 12 | 21 | 16 | 24 | 380% |
| *TPEL* | 13 | 19 | 21 | 42 | 45 | 82 | 531% |
| *IEEE Power Electronics Magazine* | 3 | 4 | 3 | 5 | 8 | 7 | 133% |
| Total | 21 | 30 | 36 | 68 | 69 | 113 | 438% |
| *Note – Data retrieved from IEEE Xplore on 24 November 2025 using the query: ("All Metadata": AI) OR ("All Metadata": "artificial intelligence") OR ("All Metadata": "deep learning") OR ("All Metadata": "machine learning").* ||||||||

While researchers and engineers are becoming more curious about AI technology and its potential to advance power electronics, far less attention has been given to the ethical implications of its adoption. Overlooking this aspect may undermine trust in the long-term of the technological advancements and practical impact of AI in a power electronics context.

Firstly, there have been an increasing number of incidents relating to AI that have begun to cause some concern in a range of sectors. The 2025 AI Index Report released by Stanford University [1] highlighted that the number of reported AI incidents by organizations has increased around 50%, with the main reported incidents including adversarial attacks, privacy violation, model bias, performance failure, and a range of other issues. Power electronics is a technology that has vital importance in many case being an essential technology used in mission-critical systems such as electric transportation and microgrids, where failures can result in consequences that are not only costly but also potentially catastrophic [2]. This combination of risk and consequence underscores the imperative to take AI ethics into consideration, to ensure that adequate safeguards are put in place, and a rigorous approach is taken to the deployment of AI in these scenarios.

Secondly, the recognition of the wider deployment of AI is leading to a strict regulatory framework in many jurisdictions, making the consideration of ethics in AI a mandatory legal responsibility rather than an optional consideration. For example, the European Union's Artificial Intelligence Act [3] entered into force on 1 August 2024. Its obligations are phased in over several years: governance rules and requirements for general-purpose AI models start applying in August 2025, while most obligations for high-risk AI systems, including legally binding requirements on transparency, accountability, and risk management, will take effect from August 2026 onward.

Against this backdrop, this article proposes an ethical framework tailored to power electronics, structured around four pillars—Security & Safety, Explainability & Transparency, Energy Sustainability, and the Evolving Roles of Engineers, as illustrated in Figure 1, with the intention of guiding future research, development, and standardization within IEEE PELS.

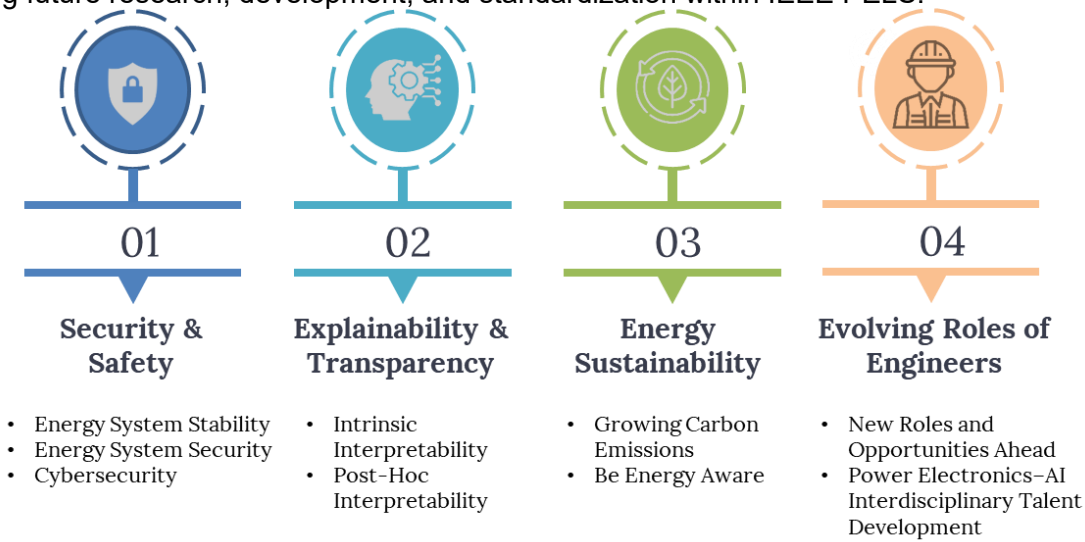

**FIG 1 Four Pillars of AI Ethics in Power Electronics**

### *AI Ethics Pillar 1: Security & Safety*

When engineers leverage AI for decision-making in power electronics, there are two key aspects of security and safety that demand attention: the **stability and security of the energy system** itself, and the **cybersecurity of the AI system**.

- **Stability And Security of the Energy System**

Focusing on energy system stability and security, every decision made by an AI algorithm carries the ethical responsibility to preserve safe and reliable operation [4]. Typical risk scenarios highlight why this is vitally important. For example, consider the scenario where AI is deployed for

Maximum Power Point Tracking (MPPT) in a photo-voltaic (PV) system. The MPPT and Inverter may encounter operating conditions never encountered in the training period, such as abrupt irradiance drops due to cloud shading. If the AI system reacts unpredictably, it may result in DC bus fluctuations, grid disturbances or even a failure condition. Similarly, in energy storage systems, power converters used to manage energy flow by either charging or discharging, enforce thermal and state-of-charge (SoC) constraints by regulating current, voltage, and power flow. An AI-driven control strategy that fails under edge conditions, such as summer peaks when both demand and battery temperature are high, could lead to overvoltage or overcharge, potentially escalating into equipment overheating, thermal runaway, or even fire.

For power electronics engineers, these ethical obligations translate into practical design workflows that incorporate validation metrics aligned with converter safety limits, reliability assessment practices, and grid-code compliance requirements, as listed in Table 2. AI models should demonstrate accuracy in matching real-world behaviour, robustness against disturbances and noise, and generalization to rare or unseen operating points. They must always respect predefined safety constraints such as current, voltage, frequency, and thermal limits, and they must deliver decisions quickly enough to meet real-time requirements. As with any engineering system, this is predicated on engineers developing adequate validation of multiple failure modes, and ensuring that testing demonstrates that the AI system can be tolerant of these events or conditions and react in a safe and reliable manner.

**TABLE 2 Technical Metrics Translated from Security and Safety Considerations**

| Technical Metrics | Notes |
|---|---|
| Accuracy | How closely AI predictions match actual values. |
| Robustness | The system's ability to maintain performance under disturbances, noise, and fault conditions (e.g., thermal spikes, grid events). |
| Generalization | AI must perform reliably in unseen or rare edge cases (e.g., extreme weather, load surge). |
| Safety Constraints | Predefined operational limits within which the AI system must operate to avoid safety risks or violations of grid codes, such as frequency / current / thermal limits, etc. |
| Latency | AI decisions must be made within strict timing requirements, especially for real-time control. |

An illustrative example of translating electrical specifications into AI engineering insights is highlighted in protection applications, with the requirements given by the IEC 61850-5 standard. Class P1 trip signals must meet an end-to-end latency requirement of less than 10 milliseconds, with even stricter limits of 3 milliseconds for Class P2/P3 [5]. If an AI model is tasked with issuing breaker trip commands, its worst-case inference time, together with communication overhead, must stay within these time limits to avoid jeopardizing equipment or personnel.

From an engineering design perspective, this calls for several practical considerations. Lightweight AI architectures are generally preferable for implementation and practicality, as they reduce inference time and make real-time deployment more feasible. Optimizing the deployment environment, for instance, through edge computing, can further minimize delays. In addition, however, robust fallback logic should always be in place, ensuring that if the AI system fails or exceeds its timing budget, a conventional protection mechanism can immediately take over. Together, these measures illustrate how abstract ethical responsibilities in AI safety are made tangible through engineering practice.

- **Cybersecurity of the AI system**

Beyond the direct impact of AI decisions on energy system stability, the cybersecurity of the AI system itself is an equally critical dimension of safety. The International Energy Agency [6] reported in 2025 that energy-sector organizations experience an average of more than 1,500 cyberattacks per week per organization. As AI becomes more tightly integrated with power electronic converters and grid control systems, it also increases the number of communication interfaces and data exchange points, creating a wider attack surface for malicious actors [7].

Consider the case of an AI system that manages a solar power plant's output by adjusting power electronic converters based on weather forecasts and energy demand. If a cyber attacker manipulates the control signals sent to the converters, the system could generate unstable voltage or frequency outputs, degrading power quality and potentially destabilizing the grid. Alternatively, a cyber attacker could tamper with the weather data that feeds the AI model, causing it to make suboptimal dispatch decisions, leading to inefficiencies or even grid instability over time, as shown in Figure 2.

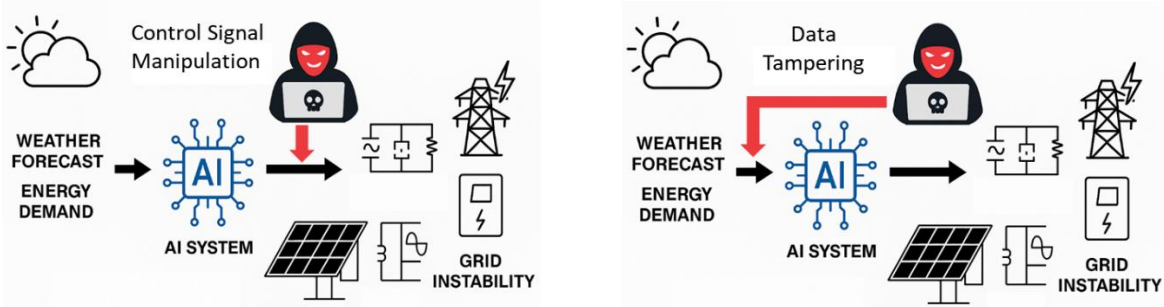

(a) Attack Scenario 1: Manipulation of control signals sent to power converters.

(b) Attack Scenario 2: Tampering with weather forecast data fed into the AI system.

**FIG 2 Two Types of Cyberattacks Targeting AI-Controlled Solar Power Plants for Energy Dispatch**

Safeguarding AI systems in this context requires a multi-layered approach, often summarized by three dimensions: model, data, and infrastructure. For models, designers should build algorithms that are robust against adversarial attacks and manipulations. Designers should also consider implementing model-integrity assessment mechanisms, such as explainable AI techniques, to ensure traceability and validation of decision-making processes. For data security, it is crucial to ensure that the information feeding the AI is accurate and tamper-free, supported by data provenance tracking to detect anomalies or corruption. At the infrastructure level, strict access controls and protection of both IT and operational technology assets are necessary to prevent unauthorized intrusion.

These measures are a stark reminder that cybersecurity is not a one-off feature but a continuous design commitment. Just as reliability engineers think in terms of fault tolerance and redundancy, power electronics engineers must design for cyber attack tolerance and recovery, ensuring that even under cyber threats, the power electronic system can continue to operate safely.

*AI Ethics Pillar 2: Interpretability & Transparency*

With AI playing a growing role in mission-critical power electronics systems, understanding and explaining model outputs is essential to build stakeholder trust. More importantly, interpretability and transparency are no longer just nice-to-have features; they have become regulatory obligations. The European Union's Artificial Intelligence Act (effective August 2025) in Article 13

[3]: "Transparency and Provision of Information to Deployers" requires that high-risk AI systems *"provide information that is relevant to explain their output."*

Interpretability can generally be pursued along two complementary paths [8], listed in Table 3:

TABLE 3 Categorization of Interpretability Approaches with Exemplary Techniques

| Categories | | | Exemplar Techniques |
|---|---|---|---|
| **Intrinsic Interpretability** | Structural Interpretability | Knowledge-Embedding | • Decision tress<br>• Fuzzy logic<br>• Physics-in-Architecture |
| | | Mathematical Interpretability | • Lipschitz-based Structural Analysis<br>• Monotonicity Constraints |
| | Learning Interpretability | Knowledge-Embedding | • Physics-in-loss<br>• Physics-in-initialization<br>• Physics-based data augmentation |
| | | Mathematical Interpretability | • Lipschitz-based Learning Stability Analysis<br>• PAC (Probably Approximately Correct) Bound |
| **Post-Hoc Interpretability** | | | • LIME (Local Interpretable Model-agnostic Explanations)<br>• SHAP (Shapley Additive Explanations) |

- **Intrinsic Interpretability**: Refers to models whose structure is inherently understandable, such as decision trees, fuzzy-logic controllers, or physics-informed neural networks. Additional mathematical guarantees such as Lipschitz-based stability analysis or monotonicity constraints help ensure that model behaviour remains predictable and physically consistent.

- **Post-hoc Interpretability**: Involves analysing complex "black-box" models after training to reveal how they make decisions. Tools such as Local Interpretable Model-agnostic Explanations (LIME) and Shapley Additive Explanation (SHAP) can identify which inputs most strongly influenced a specific prediction, thereby building trust without sacrificing model complexity.

Choosing between these two approaches often involves a trade-off between interpretability and performance [8]. Simpler models are easier to interpret intrinsically but may have lower performance. More sophisticated models can achieve higher performance but depend on post-hoc interpretability tools to be trusted in high-risk contexts.

In power electronics, the appropriate balance usually depends on the application's risk level and performance requirements. To illustrate how interpretability requirements vary with the risk profile of an application, two examples are discussed below:

- **Case 1** – AI for Power Converter Control in Mission-Critical Applications: In mission-critical applications, where AI is employed to assist in power converter control, a high level of interpretability is essential because decisions directly affect safety and reliability. In such cases, both industry stakeholders and regulatory bodies require transparent explanations of how AI-based decisions are made. Therefore, power-electronics engineers must carefully design or select models that balance interpretability and accuracy. Physics-informed AI can offer a promising approach by combining domain knowledge with data-driven learning to enhance trustworthiness.

- **Case 2** – AI for Seasonal Solar Generation Forecasting: This task involves forecasting solar generation to coordinate energy storage system operation for optimal performance. Compared with Case 1, the outcomes of these forecasts are less safety-critical, and suboptimal decisions rarely lead to severe consequences. Therefore, while interpretability remains valuable, the requirements here are typically less stringent. Black-box AI models, supplemented by post-hoc interpretability tools to analyze how inputs affect outputs, can be sufficient as long as prediction accuracy is prioritized.

Ultimately, interpretability and transparency bridge the gap between algorithmic intelligence and engineering accountability, enabling engineers to justify AI-driven decisions and maintain trust in safety-critical energy systems.

### *AI Ethics Pillar 3: Energy Sustainability*

Recent advances in AI have drawn growing attention from the power electronics community, but these gains usually come with increased computational demands. Achieving higher accuracy in many cases requires larger models and more intensive training, which not only raises financial costs but also increases carbon emissions [9]. For a field dedicated to improving power conversion efficiency and reducing energy losses, it would be counterproductive if AI developments and deployments themselves became significant energy consumers negating many previous gains. This perspective is reinforced by recent policies, such as the Energy Efficiency Directive (EED) [10] and Carbon Border Adjustment Mechanism (CBAM) [11], which introduce stricter expectations on the energy use and carbon footprint of digital and computational technologies, including AI. The goal is to ensure that AI applications in power electronics contribute to a net-positive impact on sustainability rather than undermining it.

A common methodology applied to understand the energy implications of AI is to observe the typical energy consumption breakdown across the AI development lifecycle, as shown in Figure 3. Most of the energy is consumed in three main stages:

- **Data preprocessing**: which involves cleaning, augmenting, and transforming raw data into usable formats;
- **Model training**: often the most computationally intensive step, especially for large neural networks;
- **Model inference**: where trained models are deployed to make predictions in real time or in batch operations.

Among these, model training tends to dominate energy use, as it may require thousands of GPU hours to tune model parameters and perform hyperparameter searches. However, in power electronics applications where models are deployed at scale (e.g., inverters, microgrids, battery systems), model inference energy can become significant because it runs continuously in the field.

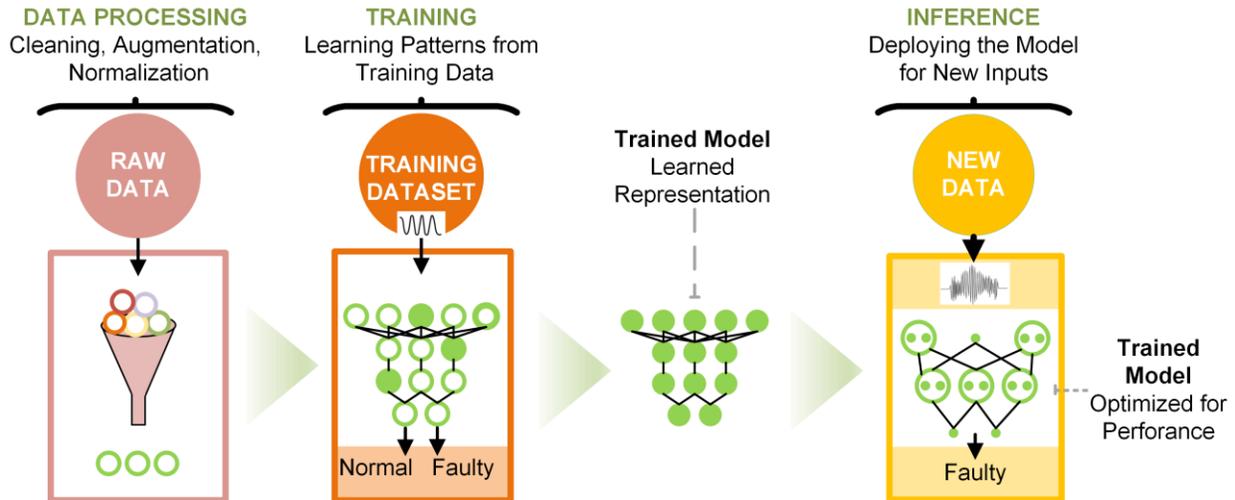
**FIG 3 Energy Usage Breakdown in AI Model Development**

To better quantify these costs, the ubiquitous Multiply–Accumulate (MAC) operation is used as a common measure of AI model complexity and energy demand as it is intrinsic to the computational cost in most AI algorithms. The MAC count serves as a hardware-agnostic proxy for the total amount of computation required by a model, allowing engineers to estimate and compare the energy footprint of different AI architectures. By incorporating MAC analysis early in the design process, power electronics engineers can make informed choices about the trade-off between model accuracy and energy sustainability.

The energy profile of large language models (LLMs) is slightly different from conventional AI workflows. Instead of training from scratch, it is more common to customize an off-the-shelf LLM using techniques such as prompt engineering, retrieval-augmented generation (RAG), or fine-tuning. While this approach avoids the massive cost of full training, it still requires energy for both customization and inference, and there is the hidden cost of training and evaluation of large data sets prior to the deployment on a specific system or design problem. The PE-GPT LLM platform for power electronics design is an illustrative example [12]. To configure PE-GPT with RAG for a simple dual-active-bridge converter modulation design case, approximately 5,500 tokens are processed to build the knowledge base, consuming about 0.005 kWh of energy. Here, a *token* refers to the smallest data unit processed by large language models. A simple use case with eight conversational rounds, totaling about 1,100 tokens, adds 0.001 kWh during one time of inference. Altogether, the one-time customization plus a single inference session requires about 0.006 kWh, roughly the energy needed to power a 60 W light bulb for six minutes. Additional rounds of interaction increase the inference energy consumption in proportion to the number of tokens processed.

These quantitative estimates not only enable power electronics practitioners to make informed design and deployment decisions but also point to research opportunities in developing energy-efficient and sustainable AI workflows tailored to the specific requirements of power electronics.

*AI Ethics Pillar 4: Evolving Roles of Engineers*

Some cutting-edge AI models have already surpassed human baselines, for instance, large language models outperform human experts in predicting neuroscience results [13], and AlphaGeometry solved 25 of 30 International Mathematical Olympiad geometry problems, performing at the level of an IMO silver medalist [14]. These illustrate that in certain narrowly defined tasks, AI has already reached or exceeded human-expert levels, raising concerning questions for power electronics about where and how AI might complement or even supplant human work. This is the same basic argument from when desktop computing began to enter the mainstream in the workplace, and yet we still require humans for most aspects of work.

What is reassuring, however surprising this may sound, is that studies have projected that AI will create more jobs rather than replace them. The Future of Jobs Report 2025 [15] by the World Economic Forum projects that, by 2030, while approximately 92 million jobs may be displaced globally due to automation and AI, around 170 million *new* jobs are expected to be created, resulting in a net positive growth of 78 million jobs. In other words, the future of work is less about mass unemployment due to the adoption of AI and more about job transformation in the world of AI.

For power electronics engineers, this means the challenge is not resisting AI but reshaping their roles to work alongside it. Ideally, in the future era of AI, rather than designing every power converter manually, engineers may increasingly act as system orchestrators, specifying design requirements, validating AI-generated solutions, and integrating them into manufacturing and operational workflows. This shift of role elevates the importance of data literacy, interpretability, and ethical decision-making, complementing traditional domain expertise.

AI also are expected to create entirely new opportunities. The upcoming EU Artificial Intelligence Act mandates transparency (Article 13) and human oversight (Article 14) [3] for high-risk AI systems, which opens doors to new professional roles, such as:

- **Documentation & Compliance Engineers**: preparing legally compliant documentation, audit trails, and user guidelines for AI-powered power systems.
- **AI Oversight Engineers**: designing systems that allow human engineers to override or correct AI behaviour safely, ensuring grid stability and equipment protection.

Both of these exemplary roles demand a rare combination of expertise: deep understanding of AI decision pipelines and energy systems, coupled with a deep knowledge of regulatory compliance to guarantee safe and lawful human-in-the-loop control. As AI adoption grows, many more roles may emerge, some of which are beyond our imagination today in 2025.

Given this rapidly changing landscape, it may be more meaningful to focus on preparing the workforce for the new roles created by AI rather than debating whether power electronics engineers will lose jobs. Thus, it is time for IEEE PELS to support future talent development by fostering interdisciplinary competence and hands-on AI engagement within the power electronics community. To achieve this, more structured educational pathways must be established. This includes the development of modernized curricula that seamlessly integrate AI concepts into traditional power electronics coursework. Dedicated training modules, micro-credentials, and laboratory experiences that blend hardware experimentation with AI algorithms will be essential to equip students and professionals alike with both theoretical understanding and practical proficiency. Besides, initiatives such as open-source community, AI campaigns, datasets, and cross-disciplinary design challenges, including the IEEE PELS open source repository hosted on GitHub to provide an index of high-quality open-sourced AI tools [16], the Magnet Challenge led by Princeton University [17], and the AI Challenge at the 2025 IEEE Design Methodology Conference led by the University of Arkansas [18], could help accelerate skill development and broaden participation. This will enable future engineers to confidently embrace the AI wave and help steer the power electronics industry toward a more intelligent, secure, and sustainable era.

## *Conclusion*

AI is rapidly advancing within the power electronics community, making it just as essential to address ethical considerations as it is to celebrate technical advancements such as accuracy. Ethics is not the sole concern of sociologists, anthropologists, or lawyers, and engineers must be part of the discussion and the implementation. Indeed, we argue that engineers should *lead* this discussion having the deepest understanding of both the technology of AI and also power electronics.

The four pillars outlined in this article provide a structural foundation for incorporating ethical responsibility into power electronics engineering practices, research and workforce development. They also create a clear pathway for IEEE PELS to contribute through recommended practices, design guidelines, interdisciplinary training resources, and future standardization efforts that enable trustworthy integration of AI into the field. By exercising proactive leadership, the community can help ensure that AI-enabled power electronics advancements are not only innovative, but also secure, transparent, energy-conscious, and clearly serving the public good.

**Acknowledgement**
The authors sincerely thank Dr. Jingxi Liu (PhD in Law, Queen Mary University of London) for her generous input and insightful legal perspective on AI regulation, which significantly informed the development of this article.

**Author Short Bio:**
- **Fanfan Lin:** Assistant Professor, Electrical Engineering, Zhejiang University-University of Illinois Urbana-Champaign Institute
- **Peter Wilon:** Professor of Electronic and Electrical Engineering, University of Bath
- **Xinze Li:** Postdoc Researcher & Lecturer, Electrical Engineering, University of Arkansas
- **Alan Mantooth:** Distinguished Professor of Electrical Engineering, the University of Arkansas